\documentclass[preprint,aps,showpacs]{revtex4}

\usepackage{makeidx}
\usepackage{graphicx}

\begin{document}

\title{A model for interevent times with long tails and multifractality in human communications: An application to financial trading} 

\author{Josep Perell\'o}
\email{josep.perello@ub.edu}
\affiliation{Departament de F\'{\i}sica Fonamental, Universitat de Barcelona,\\
Diagonal, 647, E-08028 Barcelona, Spain}
\author{Jaume Masoliver}
\email{jaume.masoliver@ub.edu}
\affiliation{Departament de F\'{\i}sica Fonamental, Universitat de Barcelona,\\
Diagonal, 647, E-08028 Barcelona, Spain}
\author{Andrzej Kasprzak}
\email{Andrzej.Kasprzak@fuw.edu.pl}
\affiliation{Faculty of Physics, University of Warsaw, Ho\.za 69, Pl-00681 Warsaw, Poland}
\author{Ryszard Kutner}
\email{Ryszard.Kutner@fuw.edu.pl}
\affiliation{Faculty of Physics, University of Warsaw, Ho\.za 69, Pl-00681 Warsaw, Poland}

\date{\today}

\begin{abstract}
Social, technological and economic time series are divided by events which are usually assumed to be random albeit with some hierarchical structure. It is well known that the interevent statistics observed in these contexts differs from the Poissonian profile by being long-tailed distributed with resting and active periods interwoven. Understanding mechanisms generating consistent statistics have therefore become a central issue. The approach we present is taken from the Continuous Time Random Walk formalism and represents an analytical alternative to models of non-trivial priority that have been recently proposed. Our analysis also goes one step further by looking at the multifractal structure of the interevent times of human decisions. We here analyze the inter-transaction time intervals of several financial markets. We observe that empirical data describes a subtle multifractal behavior. Our model explains this structure by taking the pausing-time density in the form of a superstatistics where the integral kernel quantifies the heterogeneous nature of the executed tasks. An stretched exponential kernel provides a multifractal profile valid for a certain limited range. A suggested heuristic analytical profile is capable of covering a broader region. 
\end{abstract}
\maketitle

\section{Introduction}

The dynamics of many complex systems, not only in natural sciences but in economical and social contexts as well, is usually presented in the form of time series. These series are frequently separated by random events which, in spite of their randomness, show some structure and apparent universal features~\cite{barabasi,vazquez,stanley,nakamura,barabasi2,simonsen}. During the last few years there have been endeavors to explain the sort of actions involved in interhuman communication~\cite{barabasi,vazquez2}. According to this framework, decisions are taken based on a queuing process and are aimed to be valid for a wide range of phenomena such as correspondence among people, the consecutive visits of a web portal or even transactions and trading in financial markets~\cite{barabasi2}. The main conclusion of these studies is that, in order to reproduce the empirical observations as well as to give reason of the heterogeneous nature of outgoing tasks, the timing decision has to adopt a rule of non-trivial priority. Otherwise, the implementation of, for instance, the simple rule: ``first-in-first-out'' leads to Poissonian timing between consecutive outgoing events and this seems to deviate from many empirical observations.

One convenient frame to approach these phenomena is provided by the Continuous Time Random Walk (CTRW). Within this frame one is basically concerned with the appropriate description of $\psi(t)$, the so-called pausing-time density (PTD), which gives the probability of having a certain time interval $t$ between two consecutive events. Many empirical PTD's present long-tailed profiles suggesting a self-similar hierarchy in the entire probability distribution. Following this indication some authors~\cite{barabasi2} claim that the slow decay of the PTD obeys a power-law $\psi(t)\sim t^{-\delta}$ whose exponent is almost universal in the sense that it seems to adopt only two different values $\delta=1$ and $\delta=3/2$~\cite{barabasi2}. In the next section we will present a simple approach which gives a power-law reproducing these exponents. 

Besides the PTD which doubtlessly provides maximal information on interevent statistics, the deep structure of the fractal hierarchy is perhaps more easily unveiled by looking at the $q$-moments of the interevent times instead of solely observing the PTD tails. One is thus able to answer questions such as whether the process is monofractal or multifractal and if there eventually exist different regimes depending on the value of $q$ (the order of the $q$-moment). This information obtained from data can afterwards guide us to find out the main ingredients of a more refined theoretical model for human decision dynamics. This is certainly the chief motivation of this work. 

Herein we propose an alternative framework to the existing ones --which are basically based on queuing processes-- but that it still considers the heterogeneous nature of the executed tasks. Within our approach it is possible to deal with analytical expressions, not only simulations, and we believe we provide good tools to describe the more subtle structure arisen from $q$-moments. 

The approach we propose has its roots in physics and is reminiscent of Mixture of Distributions Hypothesis in Finance that can be traced back to
the 1970s~\cite{clark}, the variational principle of energy dissipation distributions at different timescales in turbulence in the 1990s~\cite{cast}, the superstatistics and nonextensive entropy~\cite{tsallis}. In fact, the PTD $\psi(t)$ was first introduced within the CTRW model which was originally established by Montroll and Weiss~\cite{MW,weissbook,pfister}. Under this very general setting, the present development has been inspired by the work of Scher and Montroll~\cite{SM} who in 1975 proposed the so-called ``valley model'' to describe the power-law relaxation of photocurrents created in amorphous (glossy) materials. We shall use the same idea but in a completely different background.

The paper is organized as follows. In Sect. \ref{sec2} we present the fundamentals of Scher and Montroll's model and apply it to explain the emergence of long-tailed distributions in the pausing-time statistics. In Sect. \ref{sec3} we address the question of the moments of the interevent times and obtain the conditions for the multifractal behavior of such moments. In Sect. \ref{sec4} we test multifractality on large financial data sets. Conclusions are drawn in Sect. \ref{sec5} and some technical details are in the Appendix. 

\section{The Valley Model and the pausing-time Density}
\label{sec2}

Scher and Montroll's ``valley model'' proposes a conditional PTD $\psi(t|\varepsilon)$ as the starting distribution. This conditional density accounts for the probability that a given carrier is trapped during a time interval $t$ within a potential well of depth $\varepsilon$. After this time interval has elapsed the carrier jumps to another potential valley. It is next assumed that the energy $\varepsilon$ is a random variable described by a density $\rho(\varepsilon)$~\cite{SM,KKK}. We thus have a ``superstatistics'' with the unconditional pausing-time density 
$\psi(t)$ given by
\begin{equation}
\psi(t)=\int_{-\infty}^{\infty}\psi(t|\varepsilon)\rho(\varepsilon)d\varepsilon.
\label{rown:supstat}
\end{equation}
The conditional PTD is assumed to be the simple exponential (Poisson) form~\cite{SM}
\begin{equation}
\psi (t|\varepsilon )=\frac{1}{\tau (\varepsilon )}\exp \left[-\frac{t}{\tau (\varepsilon )}\right].
\label{rown:psicond}
\end{equation}
This choice is quite reasonable since for a given $\varepsilon$ the emerging statistics is homogeneous because all occurrences have the same origin and in consequence they enjoy an identical characteristic time scale $\tau(\varepsilon)$. Scher and Montroll also assume that the relationship between the random energy $\varepsilon$ and the characteristic time of the distribution is given by the simple exponential form:
\begin{equation}
\tau (\varepsilon )=\tau _0e^{\beta \varepsilon} \qquad (\tau_0,\beta>0),
\label{rown:taueps}
\end{equation}
where $\tau_0=\tau(0)$ and $\beta^{-1}$, a fundamental constant of the model, is measured in units of energy. We should note that in Scher-Montroll's approach $\beta^{-1}=K_BT$ is the thermal energy of the environment at temperature $T$ ($K_B$ is the Boltzmann constant). 

We remark at this point that the valley model is consistent with the most basic properties of a queuing process recently addressed by Vazquez et al~\cite{barabasi2}. Indeed, in that process a set of incoming messages, or tasks, arrives at random. To these messages a certain priority labeled by $\varepsilon$ is attached. The execution time of a given task with priority $\varepsilon$ is described by the conditional density $\psi(t|\varepsilon)$. In the most general setting $\varepsilon$ is also a random variable characterized by a density $\rho(\varepsilon)$. We are thus faced again with the ``superstatistics'' mentioned above since the timing of the outgoing tasks is governed by the unconditional PTD $\psi(t)$ given by Eq. (\ref{rown:supstat}). 

In the simplest case of a ``first-in-first-out'' queue the priority $\varepsilon=\mu$ has the same value for all tasks, hence $\rho(\varepsilon)=\delta(\varepsilon-\mu)$ and the unconditional PTD reads 
\begin{equation}
\psi (t)=\frac{1}{\tau(\mu)}\exp\left[-\frac{t}{\tau(\mu)}\right],
\label{poissonian}
\end{equation}
which is a Poissonian density with a single characteristic time scale (the mean time between consecutive outgoing events) 
$\tau(\mu)=\tau_0 e^{\beta\mu}$. In terms of decision theory, the situation is comparable to that of having no priority protocol at all~\cite{barabasi2}. 

\begin{figure}[htb]
\begin{center}
\includegraphics[scale=0.7]{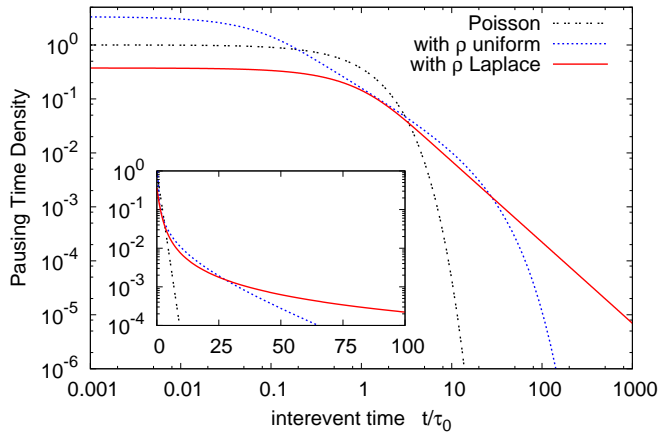}
\caption{(Color online) The pausing-time distribution for different cases as a function of the $t/\tau_0$. Dashed line provides the Poisson case as given by Eq.~(\ref{poissonian}). Dotted line is the PTD Eq.~(\ref{psi_t1_exact}) when $\rho(\varepsilon)$ is uniform ($\tau_0=1$, $\Delta=3$ $\beta=1$). Solid line represents the PTD Eq.~(\ref{exact_PTD}) when $\rho(\varepsilon)$ follows the Laplace density~(\ref{laplace}) ($\sigma=2$ and $\beta=1$. Hence the power law exponent is $3/2$, cf. Eq.~(\ref{zipf_law})).}
\label{figure:tq}
\end{center}
\end{figure}

Another particular situation would be to assign priorities in a uniform random manner with $\rho(\varepsilon)=1/(2\Delta)$ where possible values of $\varepsilon$ are constrained inside the interval $[-\Delta,\Delta]$. In this case
$$
\psi(t)=\frac{1}{2\Delta}\int_{-\Delta}^{\Delta}\psi(t|\varepsilon)d\varepsilon
$$
and from Eqs.~(\ref{rown:psicond})--(\ref{rown:taueps}) we can easily see that
\begin{equation}
\psi(t)=\frac{1}{2\Delta\beta t}\left[\exp\left(-\frac{t}{\tau_+}\right)-\exp\left(-\frac{t}{\tau_-}\right)\right],
\label{psi_t1_exact}
\end{equation}
where $\tau_{\pm}=\tau_0\exp(\pm\Delta\beta)$ (see Fig. \ref{figure:tq}). The characteristic time scale, 
$\tau_{\rm c}$, of this model (i.e., the mean time between outgoing events $\tau_{\rm c}=\langle t\rangle$) is straightly obtained using Eq. (\ref{psi_t1_exact}) and reads
$$
\tau_{\rm c}=\frac{\cosh\Delta\beta}{\Delta\beta}\tau_0.
$$

Observe that, when $\tau_+\gg\tau_-$,
\begin{equation}
\psi(t)\approx\frac{1}{2\Delta\beta t}\,\exp\left(-\frac{t}{\tau_+}\right).
\label{rown:firstclass}
\end{equation}
This truncated power-law with exponent $1$ is precisely one of the two universal classes suggested in Ref.~\cite{barabasi2} for queuing processes which emerged when the queue had a fixed length. In our model, this simple power-law arises when the random variable $\varepsilon$ is uniformly distributed. 

The second universal class proposed by Vazquez et al~\cite{barabasi2} is given by a power-law with exponent $3/2$ which appears in their simulated series when the queue is supposed to have an arbitrary length. Let us show that within our approach we can obtain arbitrary power-laws of the form $\psi(t)\sim 1/t^{\delta}$ ($\delta>1$). To this end we shall assume that the random variable $\varepsilon$ follows the Laplace distribution
\begin{equation}
\rho(\varepsilon)=\frac{1}{2\sigma}e^{-|\varepsilon|/\sigma},
\label{laplace}
\end{equation}
where $\sigma=\langle|\varepsilon|\rangle>0$. Plugging into Eq. (\ref{rown:supstat}), assuming that the conditional PTD is Poissonian (cf Eq. (\ref{rown:psicond})) and using the exponential form of $\tau(\varepsilon)$ given by Eq. (\ref{rown:taueps}), we get
$$
\psi(t)=\frac{1}{2\sigma\tau_0}\int_{-\infty}^{\infty}\exp\left\{-\beta\varepsilon-(t/\tau_0)e^{-\beta\varepsilon}-
|\varepsilon|/\sigma\right\}d\varepsilon.
$$
The integral appearing in the right hand side of this expression and that runs over the entire real line can be splitted in two integrals over the interval $(0,\infty)$ 
$$
\psi(t)=\frac{1}{2\sigma\tau_0}[I^{(+)}(t)+I^{(-)}(t)],
$$
where
$$
I^{(\pm)}(t)=\int_{0}^{\infty}\exp\left\{\pm\beta\varepsilon-
(t/\tau_0)e^{\pm\beta\varepsilon}-\varepsilon/\sigma\right\}d\varepsilon.
$$
One can easily see by simple transformations of variables that 
$$
I^{(+)}(t)=\frac{1}{\beta}\left(\frac{\tau_0}{t}\right)^{1-1/\sigma\beta}\int_{t/\tau_0}^\infty
x^{-1/\sigma\beta}e^{-x}dx=
\frac{1}{\beta}\left(\frac{\tau_0}{t}\right)^{1-1/\sigma\beta}\Gamma\left(1-1/\sigma\beta,t/\tau_0\right)
$$
and
$$
I^{(-)}(t)=\frac{1}{\beta}\left(\frac{\tau_0}{t}\right)^{1+1/\sigma\beta}\int_0^{t/\tau_0} x^{1/\sigma\beta}e^{-x}dx=
\frac{1}{\beta}\left(\frac{\tau_0}{t}\right)^{1+1/\sigma\beta}\gamma\left(1+1/\sigma\beta,t/\tau_0\right),
$$
where $\Gamma(\alpha,z)$ and $\gamma(\alpha,z)$ are incomplete gamma functions \cite{mos}. We therefore obtain the exact PTD (see Fig. \ref{figure:tq} below)
\begin{equation}
\psi(t)= \frac{1}{2\sigma\beta\tau_0}\left[\left(\frac{\tau_0}{t}\right)^{1+1/\sigma\beta}\gamma\bigl(1+1/\sigma\beta,t/\tau_0\bigr)+
\left(\frac{\tau_0}{t}\right)^{1-1/\sigma\beta}\Gamma\bigl(1-1/\sigma\beta,t/\tau_0\bigr)\right].
\label{exact_PTD}
\end{equation}

The characteristic time scale $\tau_{\rm c}$ corresponding to this density is shown to be 
(recall that $\tau_{\rm c}=\langle t\rangle$ and see Eq. (\ref{mono2}))
$$
\tau_{\rm c}=\frac{\tau_0}{1-(\beta\sigma)^2}.
$$
This characteristic time will exist as long as $\beta\sigma<1$, i.e, for the ``high-temperature phase". Thus the process has a completely different dynamics according to whether $\beta\sigma<1$ or $\beta\sigma>1$. For in the former case the system has a finite average time between consecutive outgoing events, while in the latter case, i.e., for the ``low-temperature phase", such a time ceases to exist, which means that the long term localisation for the whole observational time has sufficient probability. We can say that in this phase event statistics is dominated by rare and extreme events~\cite{kutner1,kutner2}. 

We may therefore assert that when $\sigma=\beta^{-1}$ the system undergoes a ``phase transition''. Note that for the carrier currents of Scher-Montroll's model such a phase transition will occur when the energy fluctuations, $\sigma=\langle|\varepsilon|\rangle$, equal the environment's thermal energy, $\beta^{-1}=K_BT$. 

Let us return to Eq. (\ref{exact_PTD}). For long times $t/\tau_0\gg 1$ we can write \cite{mos}
$$
\gamma\bigl(1+1/\sigma\beta,t/\tau_0\bigr)\sim\Gamma(1+1/\sigma\beta)-(t/\tau_0)^{1/\sigma\beta}e^{-t/\tau_0}
$$
and
$$
\Gamma\bigl(1-1/\sigma\beta,t/\tau_0\bigr)\sim(t/\tau_0)^{-1/\sigma\beta}e^{-t/\tau_0}.
$$
Hence, neglecting exponentially small terms, we get
\begin{equation}
\psi(t)\sim\frac{\Gamma(1+1/\sigma\beta)}{2\sigma\beta\tau_0}\left(\frac{\tau_0}{t}\right)^{1+1/\sigma\beta} \qquad(t\gg\tau_0)
\label{zipf_law}
\end{equation}
which is a general power-law of the form $1/t^\delta$ with exponent greater than $1$ (recall that $\beta$ and $\sigma$ are both positive). Therefore, the appearance in the PTD of long tails in the form of a nontruncated power-law is exclusively due to the assumption of a nonuniform distribution (in the present case the Laplace distribution Eq. (\ref{laplace})) for the random variable $\varepsilon$. Note that in deriving the power law (\ref{zipf_law}) no other power-laws had to be imposed in intermediate stages such as the in the conditional PTD, which still maintains the Poissonian form, nor in the characteristic time scale $\tau(\epsilon)$ given by Eq. (\ref{rown:taueps}). We also observe that, contrary to the uniform distribution where $\varepsilon$ is limited by the fixed length $\Delta$, there is now no upper or lower bound for the random variable $\varepsilon$. We finally remark that one of the cases analyzed in~\cite{barabasi2}, a power-law with exponent $3/2$, comes up from Eq. (\ref{zipf_law}) when $\sigma=2/\beta$. 

\section{Pausing-time moments}
\label{sec3}

The PTD $\psi(t)$ provides maximal information about the timing between successive events. There are, nonetheless, other quantities that can more easily unveil hidden but relevant features of the process such as multifractality. This is precisely the case of the pausing-time moments:
$$
\langle t^q\rangle =\int _0^{\infty }t^q\psi(t)dt.
$$
Within our framework based on a superstatistics with some weight function $\rho(\varepsilon)$, $q$-moments are written as 
\begin{equation}
\langle t^q\rangle =\int _{-\infty }^{\infty }\langle t^q|\varepsilon\rangle 
\rho(\varepsilon)d\varepsilon.
\label{tq}
\end{equation}
where $\langle t^q|\varepsilon \rangle$ are the conditional pausing-time moments defined by
$$
\langle t^q|\varepsilon \rangle =\int _0^{\infty }t^q \psi (t|\varepsilon )dt.
$$
From Eqs. (\ref{rown:psicond}) and (\ref{rown:taueps}) we get
\begin{equation}
\langle t^q|\varepsilon \rangle =\Gamma (1+q)\tau _0^q e^{q\beta\varepsilon},
\label{rown:tqcond}
\end{equation}
and the moments read
\begin{equation}
\langle t^q\rangle=\Gamma (1+q)\tau _0^q \int _{-\infty }^{\infty }e^{q\beta\varepsilon}\rho(\varepsilon)d\varepsilon.
\label{rown:psitq}
\end{equation}
We remark that exponent $q$ may be negative. However, for the Poissonian density (\ref{rown:psicond}), the conditional moment 
$\langle t^q|\varepsilon \rangle$ and, hence $\langle t^q\rangle$, will exist as long as $q>-1$. Moreover, for the moments to exist, $\rho(\varepsilon)$ must decay faster than $e^{q\beta\varepsilon}$ as $\varepsilon\rightarrow\pm\infty$.

Before proceeding ahead let us briefly recall that the timing process will show multifractality if $q$-moments behave as 
$$
\langle t^q\rangle\sim L^{f(q)},
$$
where $f(q)$ is a nonlinear function of $q$ and $L$ is some suitable scale. When $f(q)$ is linear the process is termed as monofractal \cite{bunde}. 

We will now explore the possible emergence of multifractality depending on the choice of the weight function $\rho(\varepsilon)$. The simplest assumption is $\rho(\varepsilon)=\delta(\varepsilon-\mu)$, that is, there is no heterogeneity and all the ``valleys'' have identical depth $\mu$. In such a case we readily obtain the monofractal behavior
\begin{equation}
\langle t^q\rangle=\Gamma(1+q)\tau_0^q l^q,
\label{mono}
\end{equation}
where
\begin{equation}
l\equiv e^{\beta\mu}.
\label{l}
\end{equation}

We have shown in the previous section that the uniform $\rho(\varepsilon)$ and the Laplace $\rho(\varepsilon)$ both result in long-tailed expressions for the PTD $\psi(t)$. It is thus natural to ask ourselves whether those two long-tailed distributions (cf. Eqs. (\ref{psi_t1_exact}) and (\ref{exact_PTD})) are also responsible for any multifractal behavior of the $q$-moments. 

For the uniform distribution $\rho(\varepsilon)=1/(2\Delta)$ ($-\Delta\leq\varepsilon\leq\Delta$) we have
\begin{equation}
\langle t^q\rangle=\frac{\Gamma(q)}{2\Delta\beta}\left(\tau_+^q-\tau_-^q\right),
\label{mono1}
\end{equation}
where $\tau_{\pm}=\tau_0\exp(\pm q\Delta\beta)$. Equation~(\ref{mono1}) describes approximately a monofractal situation. The genuine monofractal structure with $\tau_+^q$ emerges when $\tau_+\gg \tau_-$ or equivalently if $\Delta\beta\gg 1$. 

For the Laplace distribution, $\rho(\varepsilon)=(1/2\sigma)e^{-|\varepsilon|/\sigma}$, we have
$$
\langle t^q\rangle=
\frac{\Gamma(1+q)}{2\sigma}\int_{-\infty}^{\infty}e^{q\beta\varepsilon-|\varepsilon|/\sigma}d\varepsilon.
$$
Now $q$-moments will exist as long as $\beta q<\sigma^{-1}$. In such a case we get
\begin{equation}
\langle t^q\rangle=\frac{\Gamma(1+q)}{1-(q\beta\sigma)^2}\tau_0^q \qquad (q<(\beta\sigma)^{-1}),
\label{mono2}
\end{equation}
which is again a monofractal. 

Therefore, none of the two weight functions that result in exact long-tailed PTD's produce a multifractal structure. To get multifractality we should go one step further in the formalism and consider, for instance, a stretched exponential:
\begin{equation}
\rho(\varepsilon)=\frac{1}{2\sigma\Gamma(1+1/\alpha)}\exp\left\{-\left|\frac{\varepsilon-\mu}{\sigma}\right|^{\alpha}\right\},
\label{rown:rhoeps}
\end{equation}
($\alpha>0$ and $\sigma>0$). Substituting Eq. (\ref{rown:rhoeps}) into Eq. (\ref{rown:psitq}) we get after a simple change of variables the unconditional moment in the intermediate form
\begin{equation}
\langle t^q\rangle = \frac{\Gamma (1+q)}{2\Gamma(1+1/\alpha )}\,\tau_0^q\,l^{q}\, I(q),
\label{rown:tqnew}
\end{equation}
where $l$ is defined in Eq.~(\ref{l}) and
\begin{equation}
I(q)\equiv\int_{-\infty }^{\infty}
\exp\left(-|y|^{\alpha}+q\sigma\beta y \right)dy.
\label{rown:Iqclosed1}
\end{equation}
Looking at this integral we see at once that the convergence is assured as long as $\alpha>1$. In other words the unconditional moment $\langle t^q\rangle$ exists if $\alpha>1$ and also if $q>-1$ (see above). We can integrate 
Eq. (\ref{rown:Iqclosed1}) by means of a power series. In effect
$$
I(q)=2\int_0^{\infty}e^{-y^\alpha}\cosh(q\sigma\beta y)dy=
2\sum_{n=0}^{\infty}\frac{(q\sigma\beta)^{2n}}{(2n)!}\int_0^{\infty} y^{2n}e^{-y^{\alpha}}dy.
$$
Hence
\begin{equation}
\langle t^q\rangle=\frac{\Gamma(1+q)}{\Gamma(1/\alpha)}
\,\tau_0^q\,l^{q}\,\sum_{n=0}^{\infty}\frac{\Gamma((2n+1)/\alpha)}{(2n)!}(q\sigma\beta)^{2n},
\label{tq_series}
\end{equation}
which is an exact expression for the $q$-moment. However, due to the term $\Gamma((2n+1)/\alpha)$ appearing in the numerator, the convergence of the series in Eq. (\ref{tq_series}) is quite slow and difficult to evaluate from a numerical point of view. Nevertheless, when $\alpha=2$ ({\it i.e.,} for a Gaussian $\varepsilon$) the integral in Eq. (\ref{rown:Iqclosed1}) can be done exactly in closed form with the result:
$$
\langle t^q\rangle=\Gamma(1+q)\,\tau_0^q\,l^{q}\, e^{\sigma^2\beta^2q^2/4},
$$
which, after defining 
$$
L\equiv e^{(\sigma\beta)^2/4} 
$$
clearly shows for the Gaussian $\varepsilon$ the multifractal behavior of the $q$-moment:
\begin{equation}
\langle t^q\rangle=\Gamma(1+q)\,\tau_0^q\,l^{q}\,L^{q^2}.
\label{tq_gaussian}
\end{equation}

When $\alpha\neq 2$ the integral (\ref{rown:Iqclosed1}) cannot be done exactly and to elucidate any possible multifractal behavior of the $q$-moments we have to resort to approximations. To this end we first define in Eq. (\ref{rown:Iqclosed1}) a new integration variable $x$ by the change of scale $y=(\beta\sigma)^{1/(\alpha-1)}x$. We have
\begin{equation}
I(q)=\lambda^{1/\alpha}\int_{-\infty }^{\infty}\exp\biggl\{-\lambda\bigl[|x|^{\alpha}-qx\bigr]\biggr\}dx,
\label{Iq2}
\end{equation}
where
\begin{equation}
\lambda\equiv(\beta\sigma)^{\alpha/(\alpha-1)}.
\label{lambda}
\end{equation}
Let us suppose that the fluctuations of the random variable $\varepsilon$, represented by $\sigma$, are larger than ``the thermal energy of the environment'' $\beta^{-1}$, that is, $\beta\sigma>1$. In such a case the dimensionless parameter $\lambda$ is large and we can safely use the saddle-point approximation \cite{erderlyi} for the evaluation of $I(q)$. This is done in the Appendix with the result
\begin{equation}
I(q)\simeq A(q)\exp\left\{b|q|^{\alpha/(\alpha-1)}\right\},
\label{Iq3}
\end{equation}
where $A(q)$ is given in Eq. (\ref{a9}) of the Appendix, and
\begin{equation}
b=(\alpha-1)\left(\beta\sigma/\alpha\right)^{\alpha/(\alpha-1)}>0.
\label{b}
\end{equation}
Substituting Eq. (\ref{Iq3}) into Eq. (\ref{rown:tqnew}) we get an approximate analytical expression for the $q$-moment of the interevent time intervals. We write such an expression in a form that clearly enhances its multifractal (MF) character:
\begin{equation}
\langle t^q\rangle \simeq\Gamma (1+q) \tau _0^q l^{q}L^{|q|^{\alpha/(\alpha -1)}},
\label{rown:tqfinalnew}
\end{equation}
where, as before, we have introduced two different scales $l$ and $L$. The former $l=e^{\mu\beta}$ and given by Eq.~(\ref{l}) is related to dissipation because it depends on the dissipative term $\mu$ of the weight density (\ref{rown:rhoeps}). On the other hand, 
the scale
\begin{equation}
L\equiv e^{b}
\label{L}
\end{equation}
providing the multifractal behavior is the responsible for fluctuations since $b$ defined in Eq. (\ref{b}) depends on $\sigma$ which, in turn, is related to the variance of $\rho(\varepsilon)$; the latter given by $[\Gamma(3/\alpha)/\Gamma(1/\alpha)]\sigma^2$. Note that all $q$-moments considered above obey the normalization condition, i.e., they are equal to $1$ for $q=0$. 

The dissipative and fluctuating scales $l$ and $L$ merge into a single scale when $b=\mu\beta$. This equality means that dissipation $\mu$ and fluctuation $\sigma$ are linked by
\begin{equation}
\mu=k\sigma^{\alpha/(\alpha-1)},
\label{f-d2}
\end{equation}
where $k=(1-1/\alpha)(\beta/\alpha)^{1/(\alpha-1)}$. For $\alpha=2$ (the Gaussian case) this relation reads 
$$
\mu=(\sigma/2)^2
$$
which is the analog of the usual fluctuation-dissipation relation. This leads us to look at Eq. (\ref{f-d2}) as the fractional version of the fluctuation-dissipation theorem suitable to the present approach. 

Let us finally observe that when the fractional fluctuation-dissipation relation holds the monofractal and multifractal parts of the $q$-moment are both governed by the same scale, that is
$$
\langle t^q\rangle \simeq\Gamma (1+q) \tau _0^q L^{q+|q|^{\alpha/(\alpha -1)}}.
$$

\section{Financial database: An empirical analysis}
\label{sec4}

We shall now confront our analytical model with empirical data. We focus on moments and multifractality and leave for a future presentation extensive testing of the PTD's obtained in Sect. \ref{sec2} and their comparison with previous studies \cite{MRGS,skdr,mmw,SGM,rr,KS,MMPW,ps,jcz}. 

We have decided to apply our approach to financial markets because finance is one of the fields where large amounts of data are easily available. In particular we collect tick-by-tick data of futures contracts on several indices and also on a single stock (see Table \ref{table:data}). The assets chosen have a very diverse nature thus providing wide generality to our analysis.

\begin{table}[tb]
\caption{Empirical data specifications of the tick by tick intertransaction data used. These are futures contracts on German index (DAX), on the Dow Jones American index (DJI), on the Polish index (WIG20) and on the Foreign Exchange US Dollar-Deutsche Mark (USDM) and US Dollar-Euro (EURUS). We also add a single stock: Telefonica (TEF).}
\begin{tabular}{lccc}
\hline
\hline
Ticker & Time Period & No. of transactions \\ \hline
DAX & 2007/02/13--2007/06/14 & 4\,997\,027\\
TEF & 2006/01/02--2007/08/27 & 3\,010\,511\\
DJI & 2006/03/01--2007/08/27 & 3\,806\,980\\
WIG20 & 2006/06/19--2007/03/16 & 282\,007 \\
USDM & 1993/01/04--1997/07/31 & 1\,048\,590 \\
EURUS & 2007/08/01--2007/08/27 & 4\,176\,362\\
\hline
\hline
\end{tabular}
\label{table:data}
\end{table}

\begin{figure}[tb]
\begin{center}
\includegraphics[scale=0.7]{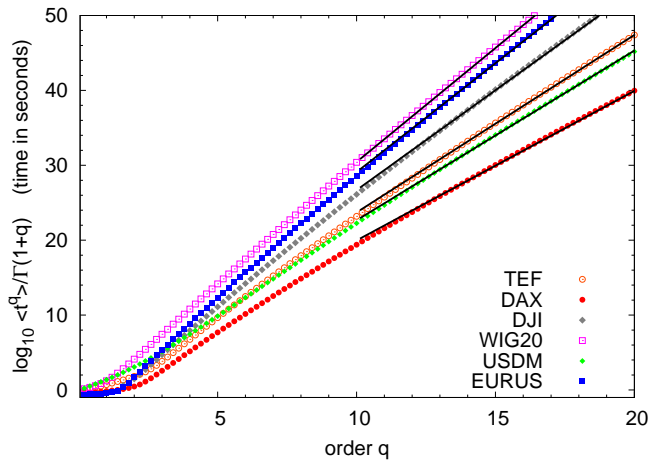}
\caption{(Color online) The normalized moment $\langle t^q\rangle /\Gamma (1+q)$ as a function of the order $q$ from six empirical financial data sets in a semi-logarithmic scale. Solid lines correspond to regressions with functions of the form $\tau^q$ for $q$ values between 10 and 20.}
\label{multifractal-long1}
\end{center}
\end{figure}

Figure~\ref{multifractal-long1} displays the empirical $q$-moments up to order $q=20$. The empirical analysis shows that all databases adopt a monofractal form for $q$ sufficiently large. To check this we define the estimate $\hat{\tau}_i$ by
\begin{equation}
{\tau}^q_i\equiv\frac{\langle t_i^q\rangle}{\Gamma (1+q)}
\label{tauhat}
\end{equation}
where $\langle t_i^q\rangle$ is the empirical $q$-moment of the database labeled by $i$. We then perform the linear regression
$$
\ln\frac{\langle t_i^q\rangle}{\Gamma (1+q)}=q\ln\tau_i,
$$
and obtain that for $10\leq q\leq 20$ the estimate ${\tau}_i$ is independent of $q$ which proves the monofractal character of the $q$-moments when $q$ is large (see Table~\ref{table:fullfit} for the specific values of ${\tau}_i$). Note that the error in estimation is very small being around 0.03\% in all markets. 

\begin{table}[tb]
\caption{Fitted parameters $\tau_i$ in seconds of the monofractal case provided by Eq.~(\ref{tauhat}) in the domain $10\leq q\leq 20$. Error in the estimation is very small.}
\begin{tabular}{lc}
\hline
\hline
$i$ dataset & $\ln\tau_i$ \\ \hline
DAX & $4.598\pm 0.004$ \\
TEF & $5.458\pm 0.004$ \\
DJI & $6.144\pm 0.005$ \\
WIG20 & $7.008\pm 0.003$ \\
USDM & $5.217\pm 0.003$ \\
EURUS & $6.697\pm 0.002$ \\
\hline
\hline
\end{tabular}
\label{table:fullfit}
\end{table}

\begin{figure}[tb]
\begin{center}
\includegraphics[scale=0.7]{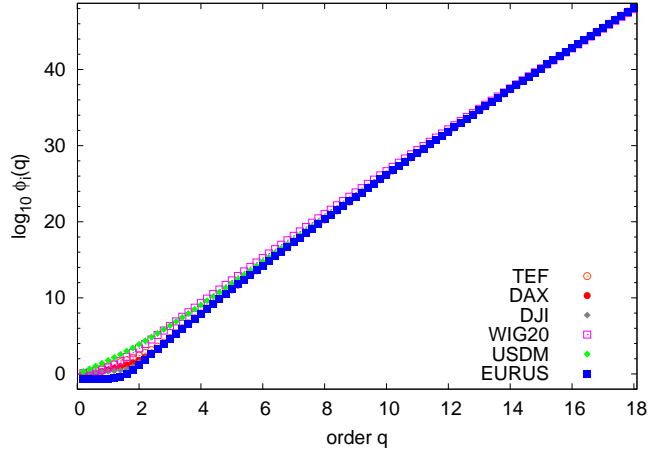}
\caption{(Color online) The ratio $\phi_i(q)$ defined in Eq. (\ref{tauhat1}) as a function of the order $q$ from six empirical financial data sets in a semi-logarithmic scale. Observe the merging of all data sets for $q>10$.}
\label{multifractal-long2}
\end{center}
\end{figure}

Let us now support the monofractal findings for $q$ large and also check the robustness of the estimate $\hat{\tau}_i$ for all data sets. To this end we evaluate the ratio
\begin{equation}
\phi_i(q)\equiv\frac{\langle t_i^q\rangle }{\Gamma (1+q)}\left(\frac{\theta}{\tau_i}\right)^q,
\label{tauhat1}
\end{equation}
where $\theta$ is an arbitrary parameter identical for all databases while $\tau_i$ is the parameter obtained from the monofractal fit when $10\leq q\leq 20$ (see Tab.~\ref{table:fullfit}). We have chosen the parameter $\theta$ to be the estimated $\tau$ for the Dow Jones Index, that is, we take $\theta=\tau_{\rm DJI}=\exp(6.144)$ (in seconds). If the monofractal hypothesis hold, $\phi_i(q)$ curves would collapse into a single straight line (in a semi-logarithmic scale). We show this analysis in Fig.~\ref{multifractal-long2} and observe that all data sets merge into a single curve for $q\geq 10$. Looking more carefully at Fig. \ref{multifractal-long2} we also observe a deviation from the monofractal behavior for $q<10$. Moreover the differences among the data sets become neatly visible for $q<5$.

\begin{figure}[tb]
\begin{center}
\includegraphics[scale=0.7]{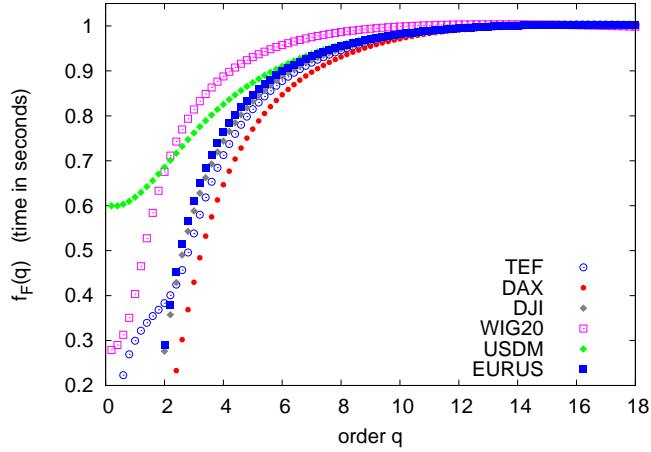}
\caption{(Color online) The function $f_{\rm F}(q)$ defined in Eq. (\ref{testm}) as a function of the order $q$ from six empirical financial data sets. Observe again the merging of all data sets for $q>10$.}
\label{figtest}
\end{center}
\end{figure}

\begin{figure}[htb]
\begin{center}
\includegraphics[scale=0.7]{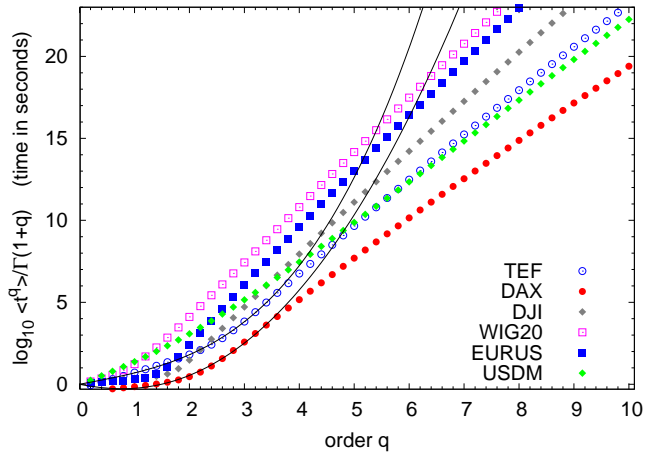}
\caption{(Color online) The emphasized plot presented in the semi-logarithmic scale, taken from Fig.~\ref{multifractal-long1}, from six empirical financial data sets for $\langle t^q\rangle /\Gamma (1+q)$. Solid lines provide two different theoretical predictions for the DAX index and Telefonica (TEF) according to multifractal formula~(\ref{rown:tqfinalnew}) with parameters given by Table~\ref{table:fullfitb}.}
\label{figure:tqshort}
\end{center}
\end{figure}

This indeed can be checked in another way than that of Fig.~\ref{multifractal-long2} by plotting the quantity
\begin{equation}
f_{\rm F}(q)\equiv\frac{1}{q\ln\tau_i}\ln(\langle t_i^q\rangle/\Gamma(1+q)).
\label{testm}
\end{equation}
If data were monofractal, the points should be close to 1 independently on the value of $q$ but as can be observed in Fig.~\ref{figtest} this is only true for $q>10$. Also observe that the points converge to 1 monotonically as the order of the moments increases. Thanks to this, the plot give us some hints on possible multifractal candidates able to fit data for the smallest orders of $q$. The description given by Eq.~(\ref{rown:tqfinalnew}) where we assumed a stretched exponential for $\rho(\varepsilon)$ appears to be a good candidate. Figure~\ref{figure:tqshort} exemplifies these abilities with the stock Telefonica and the future contracts on DAX. Table~\ref{table:fullfitb} shows the values of the estimated parameters not only for the DAX and Telefonica but also for the rest of datasets.

\begin{table}[htb]
\caption{Fitted parameters within the multifractal domain of small $q$'s. Every market has different ranges going from $q=0$ to $q=3.5$. The parameter $\alpha$ appears in general to a smaller error involved compared to other parameters.}
\begin{tabular}{lcccc}
\hline
\hline
& $q$ domain & $\alpha$ & $\ln(\tau_0)+\mu\beta$ & $b$ \\ \hline
DAX & 0-3.5 & $1.85\pm 0.13$ & $-1.5\pm 0.4$ & $0.9\pm 0.3$ \\
TEF & 0-3.5 & $1.47\pm 0.06$ & $1.45\pm 0.12$ & $0.14\pm 0.05$ \\
DJI & 0-3.0 & $1.47\pm 0.08$ & $0.34\pm 0.16$ & $0.3\pm 0.2$ \\
WIG20 & 0-2.0 & $1.65\pm 0.06$ & $1.67\pm 0.14$ & $1.19\pm 0.14$ \\
USDM & 0-3.0 & $1.89\pm 0.03$ & $2.82\pm 0.02$ & $0.32\pm 0.02$ \\
EURUS & 0-2.5 & $2.1\pm 0.2$ & $-4.6\pm 1.1$ & $3.4\pm 1.1$ \\
\hline
\hline
\end{tabular}
\label{table:fullfitb}
\end{table}

\begin{figure}[htb]
\begin{center}
\includegraphics[scale=0.7]{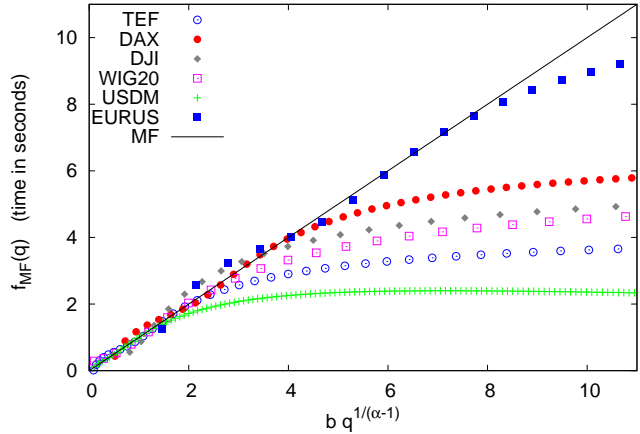}
\caption{(Color online) The function $f_{\rm MF}(q)$ defined in Eq. (\ref{test2}) as a function of the scaled moment order $bq^{1/(\alpha-1)}$ from six empirical financial data sets. Observe again the merging of all data sets for small $q$ to linear curve with slope equals 1.}
\label{figtest2}
\end{center}
\end{figure}

Summarizing we may say that empirical $q$-moments of financial interevent times clearly show multifractal behavior for (approximately) $q\leq 3.5$. In this case the analytical model expressed by Eq. (\ref{rown:tqfinalnew}),
\begin{equation}
\langle t^q\rangle \simeq\Gamma (1+q) \tau _0^q l^q L^{|q|^{\alpha/(\alpha -1)}},
\label{MF-time_2}
\end{equation}
agrees with empirical data (see Table~\ref{table:fullfitb}). On the other hand, for higher values of $q$ the data undoubtedly show a clear tendency to monofractality as has been tested in Figs.~\ref{multifractal-long2} and~\ref{figtest}. The abilities of the model~(\ref{MF-time_2}) for small $q$ can be checked as well through the function
\begin{equation}
f_{\rm MF}(q)\equiv\frac{1}{q}\ln(\langle t^q\rangle/\Gamma (1+q))-\left[\ln({\tau}_0)+\mu\beta\right].
\label{test2}
\end{equation}
We can plug into Eq.~(\ref{test2}) the parameters estimated from each data base and afterwards represent $f_{\rm MF}(q)$ as a function of $bq^{1/(\alpha-1)}$. In case we observe the merging of all databases into a straight line with slope 1 and without independent term, we could thus assert that MF model with the stretched exponential for density $\rho(\varepsilon)$ is a good candidate (cf. Eq.~(\ref{MF-time_2}) and~(\ref{test2})). This exercise is done in Fig.~\ref{figtest2}. The merging for small $q$ demonstrates the abilities of the MF model not only for the Telefonica stock and the futures on DAX but also for the rest of financial databases. 

\begin{figure}[htb]
\begin{center}
\includegraphics[scale=0.7]{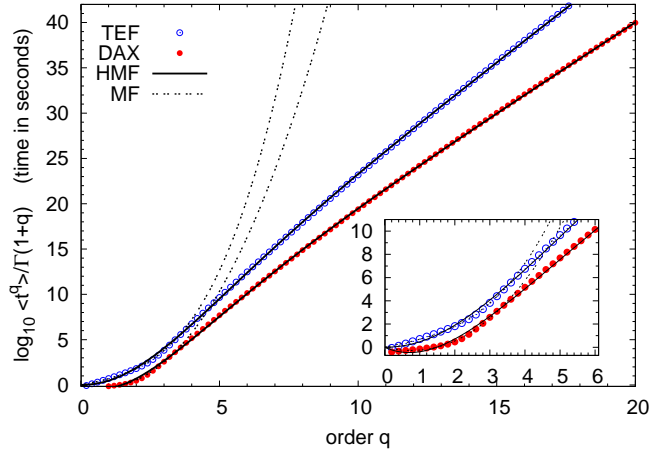}
\caption{(Color online) The normalized moment $\langle t^q\rangle /\Gamma (1+q)$ as a function of the order $q$ from Telefonica stock and DAX index in a semi-logarithmic scale. Solid lines correspond to regressions assuming the Heuristic Multifractal (HMF) formula~(\ref{rown:tqfheu}) with parameters from Table~\ref{table:tab0} while dashed lines shows again the Multifractal (MF) model with an stretched exponential superstatistics~(\ref{rown:tqfinalnew}) with parameters from Table~\ref{table:fullfit}. The inset focuses in a shorter domain of $q$ values.}
\label{figure:tqheuristic}
\end{center}
\end{figure}

\begin{table}[htb]
\caption{Set of the corresponding parameters of the heuristic extension HMF given by Eq.~(\ref{rown:tqfheu}).}
\begin{tabular}{lcccc}
\hline\hline
& $\alpha$ & $\ln\tau_0+\mu\beta$ & $b$ & $b_1$ \\ \hline
DAX & $1.91\pm 0.03$& $-3.0\pm 0.3$ & $2.5\pm 0.3$ & $0.33\pm 0.02$ \\
TEF & $1.78\pm 0.02$ & $0.1\pm 0.2$ & $1.07\pm 0.11$ & $0.20\pm 0.01$\\
DJI & $1.60\pm 0.02$ & $0.18\pm 0.12$ & $0.29\pm 0.05$ & $0.091\pm 0.012$\\
WIG20 & $1.96\pm 0.05$ & $0.5\pm 0.5$ & $3.3\pm 0.6$ & $0.50\pm 0.05$\\
USDM & $1.69\pm 0.02$ & $2.97\pm 0.06$ & $0.26\pm 0.03$& $0.115\pm 0.009$\\
EURUS & $2.21\pm 0.03$& $-9.5\pm 0.7$ & $11.7\pm 1.1$ & $0.71\pm 0.03$ \\
\hline \hline
\end{tabular}
\label{table:tab0}
\end{table}

If we want, however, to have all empirical facts in the nutshell of a single formula we should generalize Eq. (\ref{MF-time_2}) so as to include the monofractal behavior when $q$ becomes large. The requirements that such a Heuristic Multifractal (HMF) formula has to satisfy are: (i) it must obey the normalization condition (i.e., for $q=0$ it it should be equal to $1$); (ii) for small values of $q$ it must reproduce Eq. (\ref{MF-time_2}); while (iii) for larger values of $q$ the HMF formula must tend to a monofractal form. 

The heuristic formula we propose is: 
\begin{equation}
\langle t^q\rangle=\Gamma (1+q)\,\tau_0^q\,l^q\,L^{\varphi(q)},
\label{rown:tqfheu}
\end{equation}
where
\begin{equation}
\varphi(q)=\frac{1}{b_1}\left[1-\exp\left(-b_1|q|^{1/(\alpha -1)}\right)\right]|q|.
\label{exponent}
\end{equation}
Note that we have added a fourth parameter, $b_1>0$, which modifies the scale $L=\exp(b)$ by a new one $\exp(b/b_1)$ (cf. with Eq.~(\ref{rown:tqfinalnew})). Equation (\ref{rown:tqfheu}) obviously satisfies the normalization condition. Moreover, for $q$ small we have $1-\exp\left(-b_1|q|^{1/(\alpha -1)}\right)\approx b_1|q|^{1/(\alpha -1)}$ and we recover Eq. (\ref{MF-time_2}). Also for $q$ large $1-\exp\left(-b_1|q|^{1/(\alpha -1)}\right)\approx 1$ and 
Eq. (\ref{rown:tqfheu}) tends to the monofractal form:
$$
\langle t^q\rangle\simeq\Gamma(1+q)\tau_0^q\,l^q\,L^{|q|/b_1.}
$$

Figure~\ref{figure:tqheuristic} shows (solid curves) how the HMF formula fits the DAX and Telefonica empirical data on the whole range of values of $q$. This is additionally confirmed for the shorter range by the zoom provided by the inset graph. For $q<2$ the predictions of MF and HMF formulas cannot be distinguished. The value of parameters of the HMF model obtained by the fit is given for comparison in Table~\ref{table:tab0}. 

\begin{figure}[tb]
\begin{center}
\includegraphics[scale=0.7]{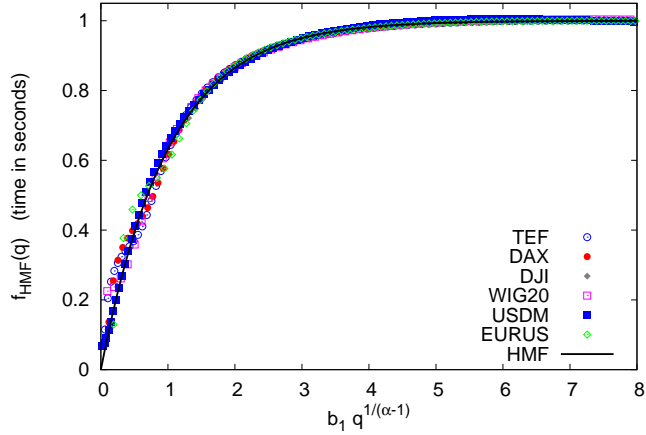}
\caption{(Color online) The function $f_{\rm HMF}(q)$ defined in Eq.~(\ref{test3}) as a function of the order $b_1q^{1/(\alpha-1)}$ from six empirical financial data sets using parameters of Table~\ref{table:tab0}. We finally observe the merging of all data sets for $0<q<20$ along a curve of the form $1-\exp(-x)$ where $x=b_1q^{1/(\alpha-1)}$.}
\label{fig:test3}
\end{center}
\end{figure}

\begin{figure}[tb]
\begin{center}
\includegraphics[scale=0.7]{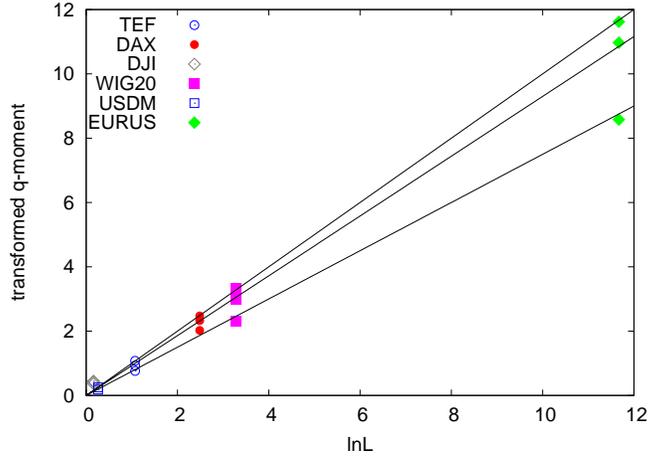}
\caption{(Color online) The transformed $q$-moment given by Eq.~(\ref{transf}) as a function of $\ln L=b$ when $q=$2 (bottom line), $q=5$ and $q=15$ (line above) and for six empirical financial data. Solid lines verify linear dependence across different markets.}
\label{fig:test3b}
\end{center}
\end{figure}

The predictions of formula~(\ref{rown:tqfheu}) have been tested on the available data sets with satisfactory results by plotting the function
\begin{equation}
f_{\rm HMF}(q)\equiv\frac{b_1}{b}\left[\frac{1}{q}\ln(\langle t^q\rangle/\Gamma (1+q))-\ln{\tau}_0-\mu\beta\right].
\label{test3}
\end{equation}
versus $b_1q^{1/(\alpha-1)}$. If the HMF hold, we then would be able to see a merging of all data sets along the curve $1-\exp(-x)$ where $x=b_1q^{1/(\alpha-1)}$ (cf. Eqs.~(\ref{rown:tqfheu}) and~(\ref{test3})). The model also allows for studying explicitly the dependence on $\ln L=b$ of transformed $q$-moment
\begin{equation}
b_1\left[\frac{1}{q}\ln(\langle t^q\rangle/\Gamma (1+q))-\ln{\tau}_0-\mu\beta\right].
\label{transf}
\end{equation}
For doing this we escale the $q$ values in each market $i$ as
\begin{equation}
{\hat q}_i=\left(b_1/b_1^i\right)^{\alpha_i-1} \, q^{(\alpha_i-1)/(\alpha-1)}
\label{scaled}
\end{equation}
where $\alpha_i$ and $b_1^i$ are these estimated parameters from market $i$ as shown in Table~\ref{table:tab0}. The unlabeled parameters $b_1$ and $\alpha$ concerns a reference market which again corresponds to the DJI futures. If the model hold, there would be a linear dependence between the transformed $q$-moments~(\ref{transf}) and $\ln L=b$ across the different markets since the HMF model~(\ref{exponent}) is invariant across the markets through
$$
f_{\rm HMF}(q)=f_{\rm HMF}(\hat{q}_i),
$$
and where $f_{\rm HMF}(q)$ would be its slope. Figures~\ref{fig:test3} and~\ref{fig:test3b} show the satisfactory results that supports the validity of the Heuristic Multifractal formula.

\begin{figure}[tb]
\begin{center}
\includegraphics[scale=0.7]{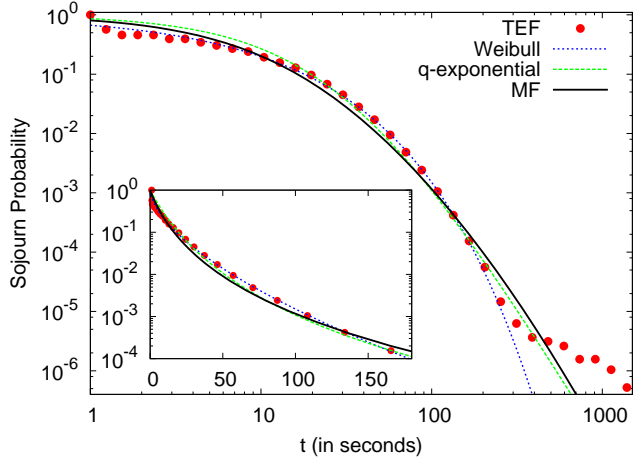}
\caption{(Color online) The sojourn probability $\Psi(t)$ of the Telefonica (TEF) stock. Solid line provides the numerical computation of $\Psi(t)$ for the MF model as given by Eq.~(\ref{Psi}) with parameters $b=0.12$, $\alpha=1.6$, and $\ln(\tau_0)+\mu\beta=1.7$. Dashed and dotted lines are respectively the fits with the q-exponential (with $m=0.94\pm 0.07$ and q$=1.3\pm 0.2$) and the Weibull probabilities (with $a=1.53\pm 0.02$ and $c=0.459\pm 0.004$).}
\label{figure:sojourn}
\end{center}
\end{figure}

We finally mention that the problem of the explicit forms of conditional PTD $\psi(t|\varepsilon)$ and distribution $\rho(\varepsilon)$ which gives the heuristic formula (\ref{rown:tqfheu}) according to relation (\ref{tq}) is still a challenge. We can otherwise check the soundness and self-consistency of our multifractal approach by looking at its sojourn probability (i.e, the decumulative probability)
\begin{equation}
\Psi(t)=\int_t^\infty \int_{-\infty}^{\infty}\psi(t'|\varepsilon)\rho(\varepsilon) dt'd\varepsilon
\label{sp}
\end{equation}
and compare it with empirical data. We take the sojourn probability instead of the pausing time distribution because verification with empirical data is firmer. Recall that the MF model takes the Poisson density $\psi(t|\varepsilon)$ provided by Eq.~(\ref{rown:psicond}) while $\rho(\varepsilon)$ obeys a stretched exponential as given in Eq.~(\ref{rown:rhoeps}). Substituting these densities into Eq.~(\ref{sp}) yields
\begin{equation}
\Psi_{\rm MF}(t)=\frac{1}{2\Gamma(1+1/\alpha)}\int_{-\infty}^{\infty} \exp\left(-|x|^\alpha-\frac{t}{\tau_0 e^{\mu\beta}}e^{-\beta\sigma x}\right) dx.
\label{Psi}
\end{equation}
The expression needs to be numerically evaluated and for doing this we have taken the parameters of Telefonica given in Table~\ref{table:fullfitb} and slightly modify them to improve the fit with empirical data. Solid line in Fig.~\ref{figure:sojourn} shows the resulting curve and it is there compared with the empirical sojourn probability of Telefonica.

The empirical analysis on the pausing time density and the sojourn probability in financial data has been extensively studied during the last few years~\cite{mmw,SGM,MMPW,ps,jcz}. Some recent papers argue that $\Psi(t)$ can be described properly by the Tsallis q-exponential~\cite{ps,jcz}
$$
\Psi_{\rm q}(t)=\frac{1}{\left[1+m({\rm q}-1)t\right]^{1/(q-1)}}
$$
with q$>1$ or the Weibull distribution~\cite{ps,jcz}
$$
\Psi_{\rm W}(t)=\exp\left(-at^c\right).
$$
These candidates are also represented in Fig.~\ref{figure:sojourn} and the quality of their fits are comparable to that of our MF model. A more accurate study among the differences and similarities of the MF model (and eventually the HMF) between both the theoretical and empirical distributions is certainly necessary. However, we leave a more complete study of the sojourn probability for a future work.

\section{Summary and conclusions}
\label{sec5}

In this work we have extended the original CTRW formalism, within the frame of Scher-Montroll's valley model, to furnish an analytical treatment for the statistics of interevent times. The model developed has been tested to financial time series, although the analysis is applicable to the broader area of interhuman communications.

The approach presented consists in obtaining the PTD $\psi(t)$ and the $q$-moments $\langle t^q\rangle$ of the interevent time intervals through a random variable $\varepsilon$ described by a probability density $\rho(\varepsilon)$. The nature of this hidden variable depends on the problem at hand. In the original work of Scher and Montroll $\varepsilon$ represented the depth of the potential well where carriers were trapped. In other contexts, such as queuing processes, $\varepsilon$ may represent the priority assigned to an incoming task and for financial markets we are exploring the possibility that $\varepsilon$ would be related with transaction volumes, market depth or bid-ask spread \cite{gopi,lillo,gillemot}. 

Whatever the case, the overall approach assumes an expression for the conditional PTD $\psi(t|\varepsilon)$ governing the timing of incoming events (charged carriers, messages, news, etc.). If these incoming events are supposed to arrive at random the natural choice for the conditional PTD is the Poisson distribution as given in Eq. (\ref{rown:psicond}). A second assumption is that for a given $\varepsilon$ the mean time between consecutive events, $\tau(\varepsilon)$, depends on the hidden variable $\varepsilon$ through the simple exponential form expressed by Eq. (\ref{rown:taueps}). Finally, in terms of the probability distribution of $\varepsilon$ the unconditional PTD and the $q$-moments (which both refer to executed tasks or outcoming events) are respectively given by Eqs. (\ref{rown:supstat}) and (\ref{tq}).

With these simple ingredients we have been able to obtain long-tailed PTD's and multifractal $q$-moments. Thus, for instance, for a Laplace density $\rho(\varepsilon)$ we have 
$$
\psi(t)\sim \frac{1}{t^\delta}, \qquad (\delta>1),
$$
which agrees with many empirical observations of diverse phenomena from queuing theory \cite{barabasi2} to finance \cite{MMPW}.

Regarding moments the choice of a stretched exponential as the probability density for $\varepsilon$ leads to a multifractal behavior of the form 
$$
\langle t^q\rangle \sim L^{|q|^\eta}, \qquad (\eta>1),
$$
where $L$ is a conveniently chosen scale. 

We have tested the multifractal behavior of intertransaction times on large financial sets of tick-by-tick data (see \cite{bacry,calvet,matteo} for multifractal analyses in other financial settings). The overall conclusion is that $q$-moments are multifractal for small values of $q$ ($q\leq 5$), while for larger orders $\langle t^q\rangle$ becomes monofractal. A more refined but heuristic analytical formula has also been proposed which fits the whole range of empirical $q$-moments. Nevertheless, the problem of the explicit forms for both the conditional PTD 
$\psi(t|\varepsilon)$ and the density $\rho(\varepsilon)$ resulting in the heuristic expression is still a challenge.

Let us finish by noting that in some places around the paper we have highlighted some thermodynamic similarities in our method. In fact the multifractal approach we have herein developed is feasible of a thermodynamic interpretation \cite{bunde,thermo}. We will develop this analogy in a future work.

\acknowledgments 
JP and JM acknowledge partial financial support from Direcci\'on General de Investigaci\'on under contract No. FIS2006-05204.

\appendix

\section{Approximate evaluation of $I(q)$}
\label{section:Iq}

We want to evaluate the integral (\ref{Iq2}) which we write in the form
\begin{equation}
I(q)=\lambda^{1/\alpha}\int_{-\infty}^{\infty}e^{-\lambda h(x)}dx,
\label{a1}
\end{equation}
where 
\begin{equation}
h(x)=|x|^{\alpha}-qx,
\label{a2}
\end{equation}
$(\alpha>1)$. For $\lambda$ large we can employ the saddle-point approximation or Laplace's method \cite{erderlyi}. Expanding $h(x)=h(x_0)+(1/2)h''(x_0)(x-x_0)^2+{\rm O}[(x-x_0)^3]$ and performing the resulting Gaussian integral we obtain
\begin{equation}
I(q)=\lambda^{1/\alpha}\left[\frac{2\pi}{\lambda h''(x_0)}\right]^{1/2}e^{-[\lambda h(x_0)+{\rm O}(\lambda^{-1/2})]},
\label{a3}
\end{equation}
where $x_0$ is the minimum of $h(x)$. That is, $x_0$ is the solution to 
$$
h'(x_0)=\alpha|x_0|^{\alpha-1}{\rm sgn}(x_0)-q=0,
$$
i.e., 
\begin{equation}
\alpha|x_0|^{\alpha-1}{\rm sgn}(x_0)=q. 
\label{a4}
\end{equation}
But $q=|q|{\rm sgn}(q)$ and we rewrite Eq. (\ref{a4}) as
$$
\frac{\alpha|x_0|^{\alpha-1}}{|q|}=\frac{{\rm sgn}(q)}{{\rm sgn}(x_0)}.
$$
Since the right hand side of this equation is positive (recall that $\alpha>0)$ then necessarily 
${\rm sgn}(x_0)= {\rm sgn}(q)$. Hence
\begin{equation}
|x_0|=\left(\frac{|q|}{\alpha}\right)^{1/(\alpha-1)},
\label{a5}
\end{equation}
and the two extremes of $h(x)$ are
$$
x_0=(q/\alpha)^{1/(\alpha-1)}
$$
when $q>0$, and
$$
x_0=-(-q/\alpha)^{1/(\alpha-1)}
$$
when $-1<q<0$. 

Taking into account that the second derivative $h''(x_0)=\alpha(\alpha-1)|x_0|^{\alpha-2}$, i.e.,
\begin{equation}
h''(x_0)=\alpha(\alpha-1)\left(\frac{|q|}{\alpha}\right)^{(\alpha-2)/(\alpha-1)}>0
\label{a6}
\end{equation}
is always positive we conclude that $x_0$ is indeed a minimum of $h(x)$. 

On the other hand, recalling that $x_0$ has the same sign as $q$ we can write $h(x_0)$ in the form
$$
h(x_0)=|x_0|^\alpha-|q||x_0|,
$$
and using Eq. (\ref{a5}) we have
\begin{equation}
h(x_0)=-(\alpha-1)\left(\frac{|q|}{\alpha}\right)^{\alpha/(\alpha-1)}.
\label{a7}
\end{equation}
Collecting terms we finally obtain
\begin{equation}
I(q)\simeq A(q)\exp\left\{b|q|^{\alpha/(\alpha-1)}\right\},
\label{a8}
\end{equation}
where 
\begin{equation}
A(q)=\lambda^{1/\alpha}\left[\frac{2\pi}{\lambda\alpha(\alpha-1)}\right]^{1/2}
\left(\frac{|q|}{\alpha}\right)^{(2-\alpha)/2(\alpha-1)},
\label{a9}
\end{equation}
and (see Eq. (\ref{lambda}))
\begin{equation}
b=\frac{\lambda(\alpha-1)}{\alpha^{\alpha/(\alpha-1)}}=(\alpha-1)\left(\frac{\beta\sigma}{\alpha}\right)^{\alpha/(\alpha-1)}.
\label{a10}
\end{equation}

\end{document}